\documentclass{article}
\usepackage{amsmath,latexsym,pst-tree,epsfig,a4wide}

\newtheorem{lemma}{Lemma}
\newtheorem{theorem}{Theorem}
\newtheorem{corollary}{Corollary}
\newenvironment{proof}{{\it Proof:\/}}{\hfill$\Box$\vskip 0.1in}

\title{Finding total unimodularity in optimization problems solved by linear programs\thanks{Supported by CNRS/NSF grant 17171 and ANR Alpage.}}

\author{Christoph D\"urr\thanks{CNRS and LIX, Ecole Polytechnique, Palaiseau,
  France, durr@lix.polytechnique.fr.} 
\and 
  Mathilde Hurand\thanks{Google, 8002 Z\"urich, Switzerland,
  mathilde@google.com.}}
\date{}

\begin{document}
\maketitle

\begin{abstract}
A popular approach in combinatorial optimization is to model problems
as integer linear programs. Ideally, the relaxed linear program would
have only integer solutions, which happens for instance when the
constraint matrix is totally unimodular. Still, sometimes it is
possible to build an integer solution with the same cost from the
fractional solution. Examples are two scheduling problems
\cite{Baptiste.Schieber:A-Note-on-Scheduling,BruckerKravchenko:Time-Windows}
and the single disk prefetching/caching problem
\cite{Albers.Garg.Leonardi:Minimizing-stall}.  We show that problems
such as the three previously mentioned can be separated into two
subproblems: (1) finding an optimal feasible set of slots, and (2)
assigning the jobs or pages to the slots. It is straigthforward to
show that the latter can be solved greedily. We are able to solve the
former with a totally unimodular linear program, from which we obtain
simple combinatorial algorithms with improved worst case running time.
\end{abstract}

\section{Introduction}
In this work, we propose a specific approach to give simpler solutions to several optimization problems. Herein we considered three such optimization problems:  the first two are scheduling problems : the \textsc{Tall-Small Jobs Problem} \cite{Baptiste.Schieber:A-Note-on-Scheduling} and the \textsc{Equal Length Jobs Problem} \cite{BruckerKravchenko:Time-Windows}. The last one is about \textsc{Offline prefetching and caching to minimize stall time} \cite{Albers.Garg.Leonardi:Minimizing-stall}. In the \textsc{Tall-Small Jobs Problem}, we have $m$ machines, $n$ unit length jobs, some of which need to execute on all the machines at the same time. In the \textsc{Equal Length Jobs Problem}, jobs have a given equal length $p\geq 1$ and each job executes on a single machine. In both problems jobs have given release times and deadlines in between which they need to execute. The goal is to find a feasible schedule, and moreover, for the equal length jobs problem, a feasible schedule that minimizes total completion time of the jobs.  
The third optimization problem,  \textsc{Offline prefetching and caching to minimize stall time}  belongs to a different field: we are given a sequence of $n$ page requests and a cache of size $k$. 
We can evict a page from the cache and fetch a new page to replace it. This operation cannot be done in parallel and costs $F$ time units. When a page request is served it costs $1$ time unit, unless the page is not yet in the cache, then a stall time is generated until the corresponding fetch completes. The goal is to decide when to evict and fetch pages so as to minimize the total stall time.

Though quite different, those three problems were solved in a similar manner. Unlike previous works where the authors  transform
the solution of a relaxed integer linear program  into an integer solution, we used a new technique which simplifies the linear programs, and allows us to get directly optimal integer solutions: our approach is based on the observation that only the structure of the solution matters in the objective function, jobs and pages don't appear namely. Therefore, we completely dissociate the resolution process into two phases. First a simplified linear program can be used to find an optimal \emph{skeleton} for the solution, and it is only later that we need to worry about \emph{assigning} jobs or pages to this skeleton: for scheduling problems, the skeleton is a sequence of slots, and the assignment maps jobs to slots; for the cache problem, the skeleton is a sequence of  intervals and the assignment associates to every interval a page to evict at the beginning and a page to fetch at the end. Our skeletons are such that the  \emph{assignment} phase just comes down to running a greedy algorithm. 
Our contribution is that this strategy, where you don't compute the assignment in the linear program, leads to linear programs with very simple constraint matrices, which not only are totally unimodular, but are (the transpose of ) directed vertex adjacency matrices.

\begin{figure}[htb]
\centerline{\epsfig{file=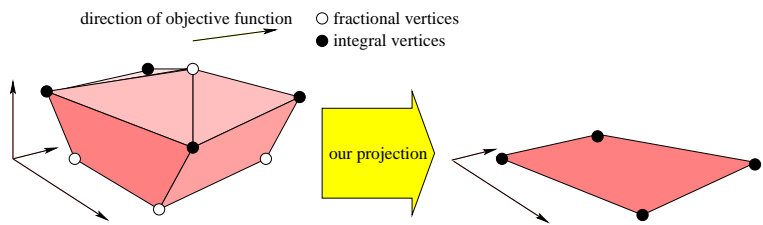,width=10cm}}
\caption{Intuition of our approach}
\label{fig:projection}
\end{figure}

This allows us to reduce our scheduling problems into a shortest path
problem and to reduce the caching problem into a min cost flow
problem.  Here is roughly our approach (see
figure~\ref{fig:projection}).  The original linear programs have the
following structure.  The polytopes described by the linear programs
have fractional vertices, but for the required objective all optimal
vertices are integral.  Now we project the solution space to a lower
dimensional space.  This is done by partitioning the variables, and
replacing every set $S$ of variables the original linear program by a
single new variable that represents the sum over $S$.  The nice thing
is that the resulting linear program is now totally unimodular, and
corresponds to shortest path or a min cost flow problem.

As a result, the tall/small job scheduling problem and the prefetch/caching problem can be solved in worst case time $O(n^3)$ improving over respectively $O(n^{10})$~\cite{Baptiste.Schieber:A-Note-on-Scheduling} and $O^{*}(n^{18})$~\cite{Albers.Buttner:Integrated-prefetching}. Implementations are available from the authors home-pages. 

\section{Scheduling equal length jobs}

We will first introduce our method on a basic scheduling problem.
We have $n$ jobs, each of the same length $p$. Every job $j\in[1,n]$ comes with an interval $[r_{j},D_{j}]$ consisting of a release time and a strict deadline. The goal is to find a schedule on $m$ parallel machines, such that each job is assigned to an execution slot consisting of a particular machine and a time interval $[s_{j},s_{j}+p) \subseteq [r_{j},D_{j}]$. In addition, all execution slots assigned to a particular machine must be disjoint.  One possible application could be frequency allocation. A network operator has a link with $m$ optical fiber strings. Users ask for allocations of a frequency band of fixed size, inside the large frequency band that the particular user devices can handle. The goal is to find an assignment which satisfies all users. In addition we want to find the solution (if it exists) that minimizes the total completion time of the jobs. 
In the standard Graham notation, this problem is called $P|r_{j};p_{j}=p;D_{j}|\sum C_{j}$.

Simons~\cite{simons78} give a complicated \emph{greedy-backtrack} algorithm running in time $O(n^{3}\log \log n)$, and later improved to $O(mn^{2})$ \cite{SimonsWarmuth:A-Fast-Algorithm}. Recently Brucker and Krav\-chen\-ko \cite{BruckerKravchenko:Time-Windows} gave another algorithm for it, using a completely different approach. While their algorithm has worse complexity it is interesting because of a generalization which permits to solve an open problem, namely minimizing the weighted total completion time, where jobs are given priority weights.

%

A generalization of the feasibility problem is to find a maximal set of jobs, which can all be scheduled between their release times and deadlines. This problem is still open. Even the more general problem, when jobs come with a weight, and the goal is to find a maximal weighted feasible job set,  is not known to be NP-hard. 

\subsection{Previous work}

First we observe that without loss of generality we can restrict ourselves to schedules where each execution slot starts at some release time plus a multiple of $p$, simply by shifting each slot as much to the beginning as possible. Let ${\cal T}=\{ r_{i}+(a-1)p : 1\le i,a \le n\}$ be this set of time points. 
And finally for a fixed schedule, if we number the execution slots from left to right, we can always reassign the $j$-th slot to the machine $(j\bmod m)+1$. This way we don't need to take care of which machines the slots are assigned to, as long as there are at most $m$ slots starting in every time interval of size $p$, which ensures that slots don't overlap on a particular machine.
The linear program of \cite{BruckerKravchenko:Time-Windows} has a variable $x_{jt}$ for each job $j$ and time $t\in\cal T$, with the meaning that $x_{jt}=1$ if job $j$ is executed in the slot $[t,t+p)$. Then the program is to minimize $\sum_{jt} (t+p) x_{jt}$ subject to
\begin{align*}
  \forall j \in [1,n] :\:& 
   	 \sum_{t\in\cal T} x_{jt} = 1 : & \text{\textbf{(every job cmpl.)}}	
 \\ 
\forall j \in [1,n], \forall t\in {\cal T} \backslash [r_{j},D_{j}-p] 
   	:\:&  x_{jt}=0 & \text{\textbf{(allowed interval)}}
 \\					
   	\forall s\in {\cal T}
   	:\:&  \sum_{s\le t < s+p} \: \sum_{j \in [1,n]} x_{jt} \le m & \text{ \textbf{(no overlapping)}}
\end{align*}

It is quite clear that there is an integer solution to this linear program if and only if there is a feasible schedule.
While this linear program is not totally unimodular, 
the authors of \cite{BruckerKravchenko:Time-Windows} were
still able to round the fractional solution into an integer solution of the same cost. 

\subsection{Relaxing the linear program}

The linear program above computes not only the time slots of the schedule, but also the assignment of jobs to slots. However once we are given the \emph{skeleton} of a schedule, meaning a set of time slots, it is always possible to assign the jobs greedily in EDD fashion: assign to every slot the job with smallest deadline among the available jobs. We release the linear program from the job assignment, in order to obtain a simpler linear program which only computes a feasible skeleton.

We proceed in several steps. First we weaken equation (\emph{every job completes}) into the inequality $\sum_{t\in\cal T} x_{jt} \geq 1$. Then combining this new constraint with (\emph{allowed interval}) leads to
\begin{equation}					\label{bk:completeAllowed}
	  	\forall j \in [1,n] :\:
			\sum _{t\in [r_{j},D_{j}-p]} 
   	:\: x_{jt} \geq 1.
\end{equation}
Now for every pair $s,t\in{\cal T}, s\leq t$ we sum (\ref{bk:completeAllowed}) over all jobs $j$ that have $[r_{j},D_{j}-p] \subseteq [s,t]$, upper-bounding the left hand side we obtain
\begin{equation}					\label{bk:HallDetail}
	\forall s,t\in{\cal T}, s\leq t :\: 
		\sum_{s'\in[s,t]}\:\sum_{j} x_{js'} \geq | \{ i: [r_{i},D_{i}-p] \subseteq [s,t] \} |.
\end{equation}

The  constraints are clearly necessary, and we will show later they are also sufficient to get the optimal solutions. We reduce the number of variables and group $\sum_{j} x_{jt}$ by setting $y_{t} := \sum_{s\leq t} \sum_{j} x_{jt}$. Now $y_{t}$ represents the total number of slots up to time $t$.
To simplify notations we introduce an additional time point $t_{0} < \min {\cal T}$, and set ${\cal T'}={\cal T}\cup\{t_{0} \}$.
For any time $t>t_{0}$, we define the functions $\text{round}(t):=\max\{ s\in{\cal T'}: s \le t\}$ and  $\text{prec}(t):=\max\{ s\in{\cal T'}: s < t\}$.

\begin{quote}
   minimize $\sum_{t\in\cal T} ( t+p ) (y_{t}-y_{\text{prec}(t)})$\\
   subject to\vspace{-1em}
   \begin{align*}
   y_{t_0}=0,\: y_{max{\cal T}} -y_{t_0} \leq n\\ 
   \forall t \in {\cal T}, s=\text{prec}(t)
   	:\:& 
	y_{s} - y_{t}\le 0 & \text{\textbf{(order)}}\\
   \forall s\in {\cal T}, t=\text{round}( s+p ) 
   	:\:& 
	y_{t} - y_{s} \le m& \text{\textbf{(load)}}\\
   \forall i,j\in[1,n], s=\text{prec}(r_{i}), t=\text{round}( D_{j}-p ), s\leq t
    	:\:&
	y_{t}-y_{s} \ge c_{ij}, & \text{\textbf{(incl.)}}
   \end{align*}
   where $c_{ij}:= | \{ k: [r_{k},D_{k}] \subseteq [r_{i},D_{j}] \} | $ is the number of jobs which have to be executed in the interval $[r_{i},D_{j}]$.
\end{quote}
The two first inequalities force $y_t$ at the first and last time
steps. In fact they are not necessary, but simplify the proof.  The
\emph{order} inequalities ensure that $(y_{t})$ is a non decreasing
sequence. The \emph{load} inequalities verify that there are never
more than $m$ slots overlapping, and the \emph{inclusion}
inequalities, are there to ensure that there is a feasible mapping
from jobs to slots, as we show next.  Except the equality $y_{t_0}=0$,
the linear program has in every constraint exactly two variables, and
with respective coefficients $+1$ and $-1$. So the dual of the
constraint matrix is the incidence matrix of a directed graph, which
means the constraint matrix is totally unimodular. Now this property
is preserved when adding a row with a single $+1$ entry, corresponding
to $y_{t_0}=0$.  Therefore our linear program's constraints are in the
from $A y \leq b$ with $A$ totally unimodular and $b$ integer. This
means that if the linear program has a solution then there is an
optimal integer solution.

Let $(y_{t})$ be an optimal integer solution to this linear
program. It indeed defines the skeleton of a solution: at each time
$t\in \cal T$ there will be $y_t - y_{prec(t)}$ slots available for
scheduling jobs.  The standard greedy job-slot assignment, is defined
as scheduling at every slot the job with the smallest deadline among
the jobs that are not yet completed and already released.
\begin{lemma}										\label{lem:valid-lp}
The greedy assignment produces a valid schedule.
\end{lemma}
\begin{proof} 
We can notice that according to the second condition and the \emph{inclusion} condition on $[t_0, max{\cal T}]$, $y_{max{\cal T}}= n$.
  We define $V$ to be the multiset of time slots, such that slot $[t,t+p]$ is contained $y_t - y_{prec(t)}$ times. Therefore $|V|=n$.  As mentioned in the previous section, by the \emph{load} inequality, the slots can be assigned to machines without overlapping. So, it only remains to show that there exist assignments of jobs to slots, which respect release times and deadlines, and then that the greedy assignment is one of them. 
  
  Let $U$ be the set of $n$ jobs, and $G(U,V,E)$ a bipartite graph
  where $E$ contains all edges between a job $j$ and a slot $[t,t+p]$
  if $[t,t+p]\in [r_{j}, D_{j}]$. We have to show that this graph has
  an injection from $U$ to $V$, and will use Hall's theorem for this,
  see \cite{Hall:Matching} or \cite{Brucker01Book}.
   
  For a set of jobs $S$, we denote the neighboring slots $\partial S$, as the set of all slots $t$ such that there is a job $j\in S$ with $(j,t)\in E$. We need to show that for every set $S$, $|S|\le |\partial S|$, which by Hall's theorem, characterizes the existence of an injection. Let $S$ be a set of jobs. Suppose $S$ can be partitioned into $S_{1}\cup S_{2}$ such that for any jobs $i\in S_{1}$ and $j\in S_{2}$ the intervals $[r_{i},D_{i}-p]$ and $[r_{j},D_{j}-p]$ are disjoint. Then clearly $\partial S$ is the disjoint union of $\partial S_{1}$ and $\partial S_{2}$. Therefore we can without loss of generality assume that $\bigcup_{j\in S}[r_{j},D_{j}]$ is a unique interval $[r_{i},D_{j}]$, for $i=\text{argmin}_{i\in S}r_{i}$ and $j=\text{argmax}_{j\in S}D_{j}$. 
  Then $|S|\le c_{ij}$. Also the number of slots in the interval $[r_{i},D_{j})$ is exactly $y_{t}-y_{s}$ for $s=\text{prec}(r_{i}), t=\text{round}( D_{j}-p )$.
  From the \emph{inclusion} inequality we get the required inequality and we conclude that there exist a valid assignment.
  Now since $|V|=|U|=n$, the injection is in fact a bijection, and there exists at least one perfect matching from jobs to slots with respect to release times and deadlines. 

Proving that you can permute jobs in any of these matching to get the greedy matching is a quite standard in scheduling: let be two jobs $i,j$ with $D_{i}<D_{j}$, and $i$ is scheduled at some time $t$, while $j$ is scheduled at some time $s$ with $r_{i}\le s < t$. Then it is possible to exchange the jobs $i,j$ in their execution slots $[s,s+p)$ and $[t,t+p)$. By  the use of a potential function, decreasing at each exchange, it is possible to transform our schedule in a so called \emph{earliest due date schedule}. 
We conclude that since there exists at least a valid assignment, the greedy assignment is valid as well.
\end{proof}

This means that an optimal integer solution can be found with a standard linear program solve. But our linear program describes in fact the dual of a minimum cost flow problem, with uncapacitated arcs, and a single supply node, which corresponds to a shortest path problem and can be solved in time $O(NM)$, where $N$ is the number of variables and $M$ the number of constraints
\cite[p.558]{Leeuwen:HandBook-of-Theoretical}.
\begin{theorem}
 Our algorithm solves $P|r_{j};p_{j}=p;D_{j}|\sum C_{j}$ in worst case time $O(n^{4})$.
\end{theorem}
\begin{proof}
Given the instance $m,p,r_{1},\ldots,r_{n},D_{1},\ldots,D_{n}$, we construct the set $\cal T$ of $O(n^{2})$ time points. Then we compute for every pair of jobs $i,j$ the number of jobs $c_{ij}$ which need to be scheduled in $[r_{i},D_{j}]$. A naive algorithm does it in time $O(n^{3})$, which would be enough for us. However it can be solved in time $O(n^{2})$ using the following recursive formula. We assume jobs are indexed in order of release times. For convenience we set $c_{n+1,j}=0$. Then $c_{i,j}=c_{i+1,j}+1$ if $D_{i}\leq D_{j}$ and $c_{i,j}=c_{i+1,j}$ if $D_{i}>D_{j}$.

This permits to construct the graph $G$ and find in time $O(n^{4})$ the optimal solution to the linear program, if there is one. Finally we do an earliest due date assignment of the jobs to the slots defined by the solution to the linear program in time $O(n\log n)$ using a priority queue.
\end{proof}

Note that in this section we don't beat the best known algorithm for $P|r_{j};p_{j}=p;D_{j}|\sum C_{j}$  which is $O(mn^{2})$ \cite{SimonsWarmuth:A-Fast-Algorithm}. However, it allows us to introduce our technique that will be used later on. 

\section{Scheduling tall and small jobs}

\begin{figure}[htb]
\centerline{\epsfig{file=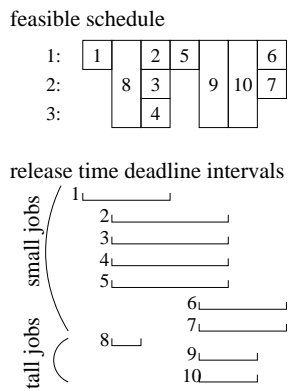,width=4cm}}
\caption{Example for 3 machines}
\label{fig:tallSmall}
\end{figure}

In a parallel machine environment, sometimes maintenance tasks are to be done which involve all machines at the same time. Think of business meetings or inventory. Formally we are given $n$ jobs of unit length $p=1$, each job $j$ comes with an integer release time and a deadline interval $[r_{j},D_{j}]$ in which it must be scheduled.
We distinguish two kinds of jobs. The first $n_{1}$ jobs are \emph{small} jobs, in the sense that they must be scheduled on one of the $m$ parallel machines, it does not matter which one. The $n_{2}=n-n_{1}$ remaining jobs are \emph{tall} jobs, in the sense that they must be scheduled on all the $m$ machines at the same time.

A time slot is an interval $[t,t+1)$ for an integer boundary $t$.
The goal is to find a \emph{feasible} schedule, where each tall job is assigned to a different time slot, and each small job is assigned to a different (machine, time slot) pair for the remaining time slots. In addition the time slot to which some job $j$ is assigned must be included in $[r_{j},D_{j}]$.

This problem has been solved by Baptiste and Schieber \cite{Baptiste.Schieber:A-Note-on-Scheduling}, with a linear program using $O(n^{2})$ variables and $O(n^{2})$ constraints. The linear program is not totally unimodular, however they manage to show that for the particular objective function it always has an integer solution. We provide a linear program using only $O(n)$ variables but still $O(n^{2})$ constraints, but whose constraint matrix is the incidence matrix of a directed graph, and can be solved in time $O(n^{3})$ with a shortest path algorithm.

Baptiste and Schieber showed that we can assume that the time interval ranges only from $1$ to $n$, otherwise the problem could easily be divided into two disjoint subproblems.
 
In a similar way than before, we will denote by $x_{t}$ the total
number of time slots assigned to tall jobs in $[1,t+1]$. For
convenience we set $x_{0}=0$.  The number of small jobs that must be
scheduled in $[s,t]$ is $k_{s,t}=|\{j:j\leq n_{1}, [r_{j},D_{j}]
\subseteq [s,t]\}|$ and the same for tall jobs is $\ell_{s,t}=|\{j:j>
n_{1}, [r_{j},D_{j}] \subseteq [s,t]\}|$. Consider the following
linear program, which does not have an objective value.

\begin{align}
							\label{ts:order}
 \forall t\in[1,n] :&\: x_{t-1}\leq x_{t}
\\
							\label{ts:onlyOne}
 \forall t\in[1,n] :&\: x_{t} - x_{t-1} \leq 1
\\
							\label{ts:serveTall}
\forall s,t\in[1,n], s\leq t:&\: x_{t-1}-x_{s-1} \geq \ell_{s,t}
\\
							\label{ts:serveSmall}
\forall s,t\in[1,n], s\leq t:&\: x_{t-1}-x_{s-1} \leq t-s -\lceil k_{s,t}/m \rceil.
\end{align}

Inequalities (\ref{ts:order}) make sure that ($x_t$)  is a non decreasing sequence, (\ref{ts:onlyOne}) that only one tall job can be scheduled per unit-length interval, (\ref{ts:serveTall}) that there are enough slots for the tall jobs and (\ref{ts:serveSmall}) that there are enough remaining slots for the small jobs.

Once again, the transpose of the constraint matrix is the adjacency matrix of an oriented graph, and the constant vector $b$ is integer. As previously, it has optimal integer solutions.  

\begin{theorem}
 Fix an instance of the tall/small scheduling problem.
 There is an integer solution to this linear program if and only if there is a feasible schedule.
\end{theorem}
\begin{proof}
It is quite obvious that fixing $(x_t)$ according to any feasible schedule will satisfy the constraints.

For the hard direction, let $(x_{t})$ be a solution to the linear program, we know it is integer. Then $x_{t}-x_{t-1}$ --- which can be $0$ or $1$ --- is the number of slots for tall jobs at time $t$. We will again use Hall's theorem to show that there is a valid assignment of the $n_{2}$ tall jobs to these slots. Inequality (\ref{ts:serveTall}) for $[s,t]=[1,n]$ forces $x_{n} \geq n_{2}$. Now let be $G(U,V,E)$ the bipartite graph, where $U$ are the $n_{2}$ tall jobs, and $V$ the $x_{n}$ slots. 
There is an edge between job $j$ and time slot $[t,t+1]$ if it is included in $[r_{j},D_{j}]$.
We have to show that for every subset $S\subseteq U$, the number of neighboring slots in $V$ is at least $|S|$. 
Let $s$ be the smallest release time among $S$ and $t$ be the largest deadline among $S$. Again it is sufficient to show this claim for connected sets $S$ in the sense that $\cup_{j\in S} [r_{j},D_{j}]=[s,t]$.
Now $|S|\leq \ell_{s,t} \leq x_{t-1}-x_{s-1}$, where the last expression is the number of slots in $[s,t]$. This completes the claim that there is a valid assignment from tall jobs to the slots. 

For the small jobs, note that $a_{s,t}:=(t-s)-(x_{t-1}-x_{s-1})$ is the number of remaining slots in $[s,t]$ which are not assigned to tall jobs, and $a_{s,t} \cdot m$ small jobs can fit in that interval. Again inequality (\ref{ts:serveSmall}) implies $k_{s,t} \leq m \cdot a_{s,t}$, and Hall's theorem shows that there is a valid assignment of small jobs to the remaining slots.
\end{proof} 

In the original paper \cite{Baptiste.Schieber:A-Note-on-Scheduling} the author gave a linear program which is solved in expected time $O(n^4)$ and worst case time $O(n^{10})$. Using the transformation into a shortest path problem allows us to improve this complexity. 

\begin{corollary}
  The tall/small scheduling problem can be solved in worst case time $O(n^{3})$.
\end{corollary}
\begin{proof}
  As in the second section, we have a linear program with $O(n)$ variables and $O(n^{2})$ constraints which can be produced in time $O(n^{2})$. We just take an arbitrary objective function in which all the variable coefficients are positive, and build the associated graph as in the previous section. Then we compute the all shortest paths from the source $x_{0}$, in time $O(n^{3})$. If this computation detects a negative cycle, then the problem has no solution. Otherwise, we get the skeleton of a solution to the problem that minimize the total completion time of the tall jobs.  Finally if there is a solution, the standard earliest due date assignment, first of tall jobs, then of small ones, produces a valid schedule in time $O(n\log n)$.
\end{proof}

Here again, a direction that we are still exploring is to find another shortest path algorithm inspired from~\cite{SimonsWarmuth:A-Fast-Algorithm}, better fitted for these specific graphs, that could improve this complexity.  

\section{Prefetching}

Caches are used to improve the memory access times. In this context the memory unit is called a \emph{page}, and is stored on a slow disk. The cache can store up to $k$ pages. Now if a page request arrives, and the page is already in the cache, it can be served immediately, otherwise it must first be fetched from the disk, and that introduces a stall time of $F$ units.  In the latter case the new page replaces some other page currently in the cache. 
The idea of \emph{prefetching} is to fetch a page even before it is requested, so as to reduce the stall time: During a fetch which evicts some page $y$ replacing it by some page $z$, other requests can be served for pages currently in the cache and different from $y$ or $z$.  In the single disk model we consider here, only a single 
fetch can occur at the same time.
The goal is, knowing in advance the complete request sequence, to come up with a prefetch schedule, which minimizes total stall time.

\begin{figure}[htbp]
\begin{center}
\epsfig{file=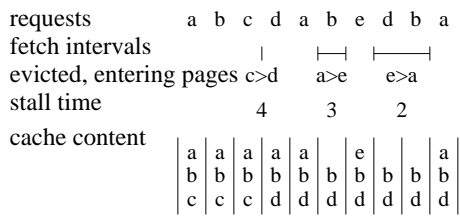,width=6cm}
\caption{An optimal prefetching for a cache of size $k=3$, and a fetch duration $F=4$.}
\label{fig:prefetch}
\end{center}
\end{figure}

While the real life problem is on-line, and has been extensively studied by Cao et al. \cite{Cao.Felten.ea:Integrated}, the offline problem has first been solved in 1998 \cite{Albers.Garg.Leonardi:Minimizing-stall}, by the use of  a linear program, for which it was shown that it always has an optimal integer solution, while not being totally unimodular. Later in 2000 \cite{Albers.Buttner:Integrated-prefetching}, a polynomial time algorithm was given modeling the problem as a multi commodity flow with some postprocessing.
Formally the problem can be defined as follows.

\paragraph{The \textsc{Offline Prefetching} problem}
The input is a page request sequence $x_{1},\ldots,x_{n}$, an initial cache set $C_{1}$, and a fetch duration $F$. Let $k=|C_{1}|$ be the cache size. A fetch is a tuple $(s,y,e,z)$, where $y,z$ are pages and $s,t\in[1,n]$ are time points with $s\leq e \leq s+F$. The meaning is 
that at time $s$, the page $y$ leaves the cache and at time $e$ the page $z$ enters the cache. Its cost, the induced stall time, is $F-(e-s)$. 
The goal is to come up with a fetch sequence minimizing the total stall time, such that two fetches intersect in at most one time point, and such that every request can be served, i.e. $\forall t\in[1,n]: x_{t}\in C_{t}$, where $C_{t}$
is the cache at time $t$ obtained from $C_{t-1}$ by evicting/fetching all the pages that had to be evicted/fetched at time $t$. To simplify notation we assume that the request sequence contains at least $k$ distinct pages, that $C_{1}$ consists of the first $k$ distinct requests, and
that at time $1$, no page has left/entered the cache yet.

Albers, Garg and Leonardi defined a linear program with a characteristic variable for every fetch interval $[s,e]$, and two additional characteristic variables for every pair ($y$, $[s,e]$) indicating whether page $y$ enters (resp. leaves) the cache at the beginning (resp. the end) of the fetch $[s,e]$. Finally they show that the linear program has always an integer solution for the considered objective function.

As observed in \cite{Albers.Garg.Leonardi:Minimizing-stall} without
loss of generality the page to be evicted at time $t$ from the cache
$C_{t-1}$ is the page, whose next request is furthest in the future or
which is never requested again. Also without loss of generality the
page to be fetched at time $t$ is the page whose next request starting
from $t$ is nearest in the future. Therefore all the information about
the fetches is in the time intervals, and we will write a linear
program which produces only the time intervals in which
evictions/fetches occur. The actual pages have to be assigned in a
post processing, in greedy manner as just mentioned.  Rather than
having a single variable for every interval and every page, we only
count how many pages entered and how many left the cache in total
since the beginning.  This leaves us with $O(n)$ instead of $O(n^2F)$
variables. We denote by $I_{t}$ (resp. $O_{t}$ ) the total number of
pages which entered (resp. left) the cache up to time $t$ included. We
get the following linear program.

\begin{quote}
minimize $FO_{n}-FI_{1}-\sum_{t=1}^{n} (O_{t}-I_{t})$,
\\
subject to\vspace{-1em}
\begin{align}
						\label{eq:order}
\forall t\in[2,n] :&\: O_{t-1} \leq O_{t}
\mbox{ and } I_{t-1} \leq I_{t}
\\
						\label{eq:overfull}
\forall t\in[1,n]:&\: O_{t} \geq I_{t}
\\
						\label{eq:overlap}
\forall t\in[1,n] :&\: O_{t} \leq I_{t}+1
\\						\label{eq:length}
\forall t\in[1,n] :&\: I_{\min\{t+F,n\}} \geq O_{t}
\\						\label{eq:serve}
\forall 1\leq s\leq t\leq n :&\: I_{t}-O_{s} \geq |\{ x_{s},x_{s+1},\ldots,x_{t} \}|  - k
\end{align}
\end{quote}

Inequalities (\ref{eq:order}) make sure that $(O_t)$ and $(I_t)$ are
non decreasing sequences, (\ref{eq:overfull}) that the cache cannot
overflow, (\ref{eq:overlap}) that two fetches don't overlap in time,
(\ref{eq:length}) that a fetch length is at most $F$ and
(\ref{eq:serve}) that there are enough fetches to serve all requests.
Note that $|\{ x_{s},x_{s+1},\ldots,x_{t} \}|$ denotes the number of
distinct page requests in $[s,t]$, so (\ref{eq:serve}) is the part of
linear program that depends on the actual problem instance. 

The optimal solution of this linear program is always integer, since
it is totally unimodular (for the same reason as in previous section:
its constraint matrix the transposed incidence matrix of a directed
graph.)

\begin{theorem}
Let $(I_{t},O_{t})$ be an optimal integer solution to the linear
program. Then there is valid fetch sequence of the same cost, which
can be built by greedy assignment.
\end{theorem}
\begin{proof}
First we observe that the cost function makes sure that $O_{n}=I_{n}$,
which ensures that all interval are eventually closed. The solution
defines $m=O_{n}$ intervals as follows. For every $j=1\ldots m$, let
$s_j$ be the smallest time such that $O_{s_j}\geq j$ and $e_j$ the
smallest time such that $I_{e_j}\geq j$. Then by (\ref{eq:overfull})
and (\ref{eq:length}) we have $s_{j}\leq e_{j}\leq s_{j}+F$. Which
means that all intervals $[s_j,e_j]$ are well defined and of length smaller or
equal than $F$. Now by (\ref{eq:overlap}), $e_j \leq s_{j+1}$
(otherwise, we would have $I_{e_j}+1 \geq O_{e_j} > O_{s_{j+1}}$ but
$I_{e_j}=j$ and $ O_{s_{j+1}}=j+1$ by definition.), and this for all
$j<m$, so the intervals do not overlap (but the ending point of one
might be the starting point of another). Moreover the objective value
of $(I_{t},O_{t})$, equals the total stall time of these intervals,
for at each time $t$, the difference $O_t$ - $I_t$ is $1$ if an
interval is currently opened and is $0$ otherwise.  It remains to
prove that the greedy assignement of pages to evict/fetch to each
interval is such that all requests are served, i.e. it remains to show
that the constraints (\ref{eq:serve}) are sufficient. We denote by
$C_{s}$ the cache obtained at time $s$, after all entrances and
evictions that occur at time $s$. We will show that the following
invariant holds in a solution of our linear program for every time
$s\in[1, n]$,
\begin{equation}					\label{eq:invariant}
	\forall t\in[s,n]: I_{t}-I_{s} \geq |\{x_{s},\ldots,x_{t}\}\backslash C_{s}|. 
\end{equation}
The invariant implies that if the number of pages requested in $[s,t]$
but not in the cache at time $s$ is $a$, then at least $a$ pages must
enter the cache somewhere in $[s+1,t]$. In particular it means for
$t=s$, that the page requested at time $s$ will be the in the cache at
that moment. The proof of (\ref{eq:invariant}) is by induction on $s$.

\paragraph{Basis case $s=1$}
Let $t_{0}$ be the greatest request time such that $x_{t_{0}}$ is not in $C_{1}$. 
Then by the assumption that initially the cache contains the first $k$ distinct requests, we have that for $t<t_{0}$, $\{x_{1},\ldots,x_{t}\} \subseteq C_{1}$. So the right hand side of (\ref{eq:invariant}) is $0$ and (\ref{eq:invariant}) holds by (\ref{eq:order}). For $t\geq t_{0}$, since the intersection of $\{x_{1},\ldots,x_{t}\}$ and $C_{1}$ is exactly $k$, the invariant holds by (\ref{eq:serve}) and $O_1 =I_1 =0$.

\paragraph{Induction case} Assume the invariant holds for some $s$. Let's show that it also holds for $s+1$. Several things can happen at time $s+1$. Pages can leave the cache and pages can enter the cache. We will do these operations step by step, transforming  $I_s$ into $I_{s+1}$ and $C_s$ into $C_{s+1}$, and show that each step preserves the invariant (\ref{eq:invariant}). 

By induction hypothesis $x_{s}\in C_{s}$, so $\{x_{s},\ldots,x_{t}\} \backslash C_{s}=\{x_{s+1},\ldots,x_{t}\} \backslash C_{s}$. Therefore, if nothing happens and no page enters or leaves the cache, then $C_{s+1}=C_s$, $I_s = I_{s+1}$ and the invariant is preserved for $s+1$.

Now we deal with the case when there is some page movement at time $s+1$, that is $I_{s+1}>I_{s}$ or $O_{s+1}>O_{s}$ or both. We artificially decompose this page movement in as many times as needed, so that at each time there is only one operation happening: a fetch or an eviction. The page movements at those intermediary times are set so as to alternatively evict and enter pages, among
the $O_{s+1}-O_{s}$ pages to evict and the $I_{s+1}-I_{s}$ pages to enter. Of course if a fetch is pending at time $s$, that is $|C_s|=k-1$, then we start with entering a new page and otherwise if the cache is full, i.e. $|C_s|=k$, we start with evicting a page. Since the number of total entrances and evictions up to some time can differ by at most one, it is possible to do so. Therefore, we need to do the induction case only in the case when a page is entering the cache or when one is leaving the cache but not both.

When page is entering the cache, we have $I_{s+1} = I_s +1$.
Let $z$ be the page entering the cache, and let $t_{0}\geq s+1$ be the next request time of $z$. Then if $t<t_{0}$, by the choice of $z$, all requests of $x_{s+1},\ldots,x_{t}$ must be in $C_{s+1}$, so the right hand side of (\ref{eq:invariant}) at time $s+1$ is $0$, and the inequality holds by (\ref{eq:order}). Now if $t\geq t_{0}$, since $z \in C_{s+1}$ but $z \not\in C_s$, the left hand side of  (\ref{eq:invariant}) at time $s+1$ has decreased  by $1$ compared to time $s$, but at the same time $I_{s+1} = I_s +1$,  so both sides of the invariant decrease by $1$ and by induction the inequality is preserved at time $s+1$.

Now consider the case when a page leaves the cache. Let  $y$ be the leaving page. Then $I_s=I_{s+1}$ and $O_{s+1}=O_s +1$. Let $t_{0}$ be the next request time of $y$ or let $t_{0}=n+1$ if $y$ is never requested again. Then if $t<t_{0}$, removing $y$ from $C_{s+1}$ does not change the right hand side of (\ref{eq:invariant}) when replacing $s$ by $s+1$. The left hand side does not change either since no page enters the cache, and the inequality is preserved. For $t\geq t_{0}$ however by the choice of the evicted page $y$, we have that $C_{s+1}\subseteq \{x_{s+1},\ldots,x_{t}\}$. So the left hand side of (\ref{eq:invariant}) at time $s+1$ is $|\{x_{s+1},\ldots,x_{t}\}|-(k-1)$, and $I_{s+1}=O_{s+1}-1$ since we have just evicted a page. Therefore, (\ref{eq:invariant}) holds by  (\ref{eq:serve}).
\end{proof}

\begin{theorem}
The offline prefetch problem can be solved in time $O(n^{3})$ if $F=O(n)$ and in time $O(n^{3}\log n)$ otherwise.
\end{theorem}
\begin{proof}
First we observe that the $O(n^n)$ different righthand sides of
(\ref{eq:serve}) can be computed in time $O(n^2 \log n)$ using dynamic
programming and a search tree, so solving the linear program is the
bottleneck of the algorithm.

The dual of the linear program is a min cost flow problem with uncapacitated arcs,
where the supply/demand $b_{i}$ of the nodes $i$ are given by the coefficients in the cost function and where the arc costs $c_{ij}$ are given by right hand sides of the inequalities, see figure~\ref{fig:graph}.

\begin{figure}[htbp]
\begin{center}
\epsfig{file=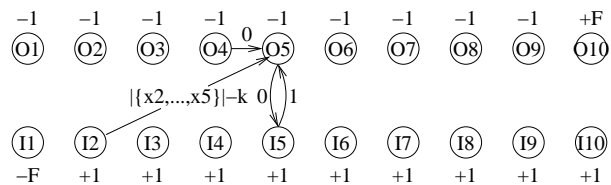,width=8cm}
\caption{Min cost flow problem instance  (not all arcs shown).}
\label{fig:graph}
\end{center}
\end{figure}

It could be solved in time $O(n^{3}\log n)$ using \cite{Orlin:A-Faster-Strongly}. To solve it in $O(n^{3})$, when $F=O(n)$, we first explode the source of supply $F-1$ into $F-1$ vertices of supply $+1$ and do the same with the sink of demand $1-F$. The new graph has only sources of supply $+1$ and sinks of demand $-1$. Clearly there is a bijection between the min cost flows of the new and the original graph. Moreover a min cost flow matches sources to sinks such that the flow between a matched source/sink pair uses a shortest path (since the arcs have unbounded capacity) and such that the total distances are minimal. To obtain this flow we first compute the distances in the graph between all source/sink pairs, in time $O(n^{3})$ using Floyd-Marshall's algorithm. Then we construct the bi-partite sources/sinks graph, where every edge is weighted with the source-sink distance in the original graph. Then a minimum weighted perfect matching can be computed in time $O(n^{3})$ provided $F\in O(n)$, using Edmond's algorithm with adapted data-structures. The optimal flow then is obtained by adding a unit flow on the shortest path between source $i$ and sink $j$, for every edge of the matching corresponding to source $i$ and sink $j$.

Finally to get an optimal solution for the primal linear program, we use the standard technique of computing a shortest path tree in the residual graph obtained from the flow \cite[chapter 9]{Ahuja.Magnanti.Orlin:Network-Flows}.
\end{proof}

\section{Conclusion}



 Further work would include trying to find other optimization problems where our technique may apply, and maybe generalize from it a general framework. We are also interested in improving the combinatorial algorithms that arise from the graph structures in the scheduling problems: indeed those graphs have, among others, the property that once the vertices drawn as points on a line, the arcs from left to right have positive weights and the ones from right to left negative.  One idea for instance is to try and extract from Simons and Warmuth's algorithm a shortest path algorithm suitable for our class of graphs.

We wish to thank Arthur Chargueraud, Philippe Baptiste, Miki Hermann and Leo Liberti for helpful comments. 
 
\bibliographystyle{plain}
\bibliography{sched}

\end{document}